\documentclass[journal]{IEEEtran}

\usepackage{tikz}
\usepackage{pgfplots}
\pgfplotsset{compat=1.17}
\usetikzlibrary{external}
\usetikzlibrary{patterns}
\usepackage{array}
\usepackage{lipsum,graphicx,multicol,multirow}
\usetikzlibrary{shapes,arrows,positioning,calc,decorations.pathreplacing}
\usetikzlibrary {arrows.meta,shapes.misc,calc}
\usetikzlibrary{graphs,graphs.standard,quotes}
\usepackage{amsmath, amssymb,amsfonts,amscd,xpatch}
\usepackage{cases}
\usepackage{mwe,comment,balance}
\usepackage{xcolor,colortbl} 
\usepackage[noadjust]{cite}

\ifCLASSOPTIONcompsoc
 \usepackage[caption=false,font=normalsize,labelfont=sf,textfont=sf]{subfig}
\else
 \usepackage[caption=false,font=footnotesize]{subfig}
\fi

\newcommand{\Exp}[1]{\mathbb{E}\left[#1\right]} 
\newcommand{\Var}[1]{\mathrm{Var}\left[#1\right]}
\newcommand{\AvgExp}[1]{\overline{\mathbb{E}}\left[#1\right]} 
\newcommand{\AvgVar}[1]{\overline{\mathrm{Var}}\left[#1\right]}
\newcommand{\TExp}[1]{\mu_{\scriptscriptstyle #1}}
\newcommand{\TVar}[1]{\sigma^2_{\scriptscriptstyle #1}}

\newcommand{\argmin}{\mathop{\mathrm{argmin}}}

\newcommand{\Prob}{\mathbb{P}}

\newcommand{\WE}{G^W}

\newcounter{lemma}

\begin{document}
\title{List-encoding CCDM: A Nonlinearity-tolerant Shaper Aided by Energy Dispersion Index}

\author{Kaiquan~Wu,~\IEEEmembership{Student Member, IEEE}, Gabriele~Liga,~\IEEEmembership{Member, IEEE}, Alireza~Sheikh,~\IEEEmembership{Member, IEEE}, Yunus~Can~G\"ultekin,~\IEEEmembership{Member, IEEE}, Frans~M.~J.~Willems,~\IEEEmembership{Life Fellow, IEEE}, and Alex~Alvarado,~\IEEEmembership{Senior Member, IEEE}
\thanks{This work is supported by the Netherlands Organisation for Scientific Research via the VIDI Grant ICONIC (project number 15685). The work of Alex Alvarado is supported by the European Research Council (ERC) under the European Union’s Horizon 2020 research and innovation programme (grant agreement No. 757791). The work of G.~Liga is supported by the EuroTechPostdoc programme under the European Union’s Horizon 2020 research and innovation programme (Marie Skłodowska-Curie grant agreement No. 754462).}
\thanks{The authors are with the Information and Communication Theory Lab, Signal Processing Systems Group, Department of Electrical Engineering, Eindhoven University of Technology, Eindhoven 5600 MB, The Netherlands (e-mails: \{k.wu, g.liga, a.sheikh, f.m.j.willems, y.c.g.gultekin, a.alvarado\}@tue.nl).}
\thanks{A. Sheikh is with imec, Holst Centre, High Tech Campus 31, 5656 AE Eindhoven, The Netherlands (email: alireza.sheikh@imec.nl).}}

\markboth{Preprint, \today}%
{Shell \MakeLowercase{\textit{et al.}}: Bare Demo of IEEEtran.cls for IEEE Journals}

\maketitle

\begin{abstract}
Recently, a metric called \emph{energy dispersion index} (EDI) was proposed to indicate the nonlinear interference (NLI) induced by correlated symbols during optical transmission. In this paper, we propose a new shaper architecture to decrease the EDI of transmitted symbols and thus, increase the signal-to-noise ratio (SNR).
We call this shaper the \emph{list-encoding} constant-composition distribution matcher (L-CCDM). L-CCDM consists of an additional EDI selecting module, which is compatible with standard probabilistic amplitude shaping (PAS) architecture. Numerical results obtained from a multi-span multi-channel system show that when compared to standard CCDM with 256-ary quadrature amplitude modulation (256QAM), the proposed architecture offers an effective SNR gain of $0.35$~dB, an achievable information rate gain of $0.22$~bit/4D-symbol, or equivalently an $8\%$ reach extension.
\end{abstract}


\IEEEpeerreviewmaketitle

\section{Introduction}

\IEEEPARstart{P}{robabilistic} amplitude shaping (PAS) \cite{bocherer2015bandwidth} has attracted wide attention in the domain of fiber optical communications. Numerous studies about PAS have been carried out in simulation \cite{fehenberger2016probabilistic,amari2019introducing}, experiments \cite{buchali2016rate,ghazisaeidi2017advanced} and field trials \cite{cho2018trans,olsson2018record}. PAS realizes probabilistic shaping by means of an amplitude shaper. One popular amplitude shaper is the constant-composition distribution matcher (CCDM) \cite{schulte2015constant}. Other well-known shapers include multiset-partition distribution matcher (MPDM) \cite{fehenberger2018multiset}, product distribution matcher (PDM) \cite{steiner2018approaching}, and enumerative sphere shaping (ESS) \cite{willems1993pragmatic,gultekin2017constellation}. 

Tailored to the additive white Gaussian noise (AWGN) channel, the aforementioned amplitude shapers were initially designed for the purpose of generating nonuniformly distributed amplitudes. However, unlike the AWGN channel, in the nonlinear fiber channel, nonlinear interactions among the transmitted symbols occur. These interactions lead to nonlinear interference (NLI) noise. As a result, the AWGN-optimal PAS usually enhances the NLI with respect to uniform signaling~\cite{gultekin2021kurtosis}. Therefore, NLI-tolerant shaping architectures, which improve power efficiency and also combat the NLI penalty, are of great interest.

The amplitude shaper determines the probability distribution as well as the \emph{temporal} structure of the transmitted symbols. For the design of NLI-tolerant amplitude shapers, one approach is to optimize the probability distribution of input symbols, while assuming the symbols to be independent identically distributed (i.i.d.). With the advent of the enhanced Gaussian noise (EGN) model, the fourth standardized moment (a.k.a. kurtosis) of transmitted symbols has been widely accepted as a metric that indicates the NLI magnitude \cite{dar2014inter,carena2014accuracy,poggiolini2015simple}. In \cite{sillekens2018simple}, an optimized probability mass function for a nonlinear fiber channel is designed taking kurtosis into account. In \cite{fehenberger2019analysisECOC,gultekin2021kurtosis}, symbol sequences having low kurtosis are generated to mitigate the NLI.

Instead of optimizing distributions, one can manipulate the \emph{temporal} structure of the symbol sequences, with the goal of reducing the NLI. This follows from the fact that the transmitted symbols that have certain temporal structures could exert great influence on the NLI \cite{agrell2014capacity,dar2014shaping,yankov2017temporal}. A straightforward approach for improving the NLI tolerance with this temporal approach was to use short shaping blocklengths \cite{geller2016shaping,fehenberger2020mitigating}. This nonlinear shaping gain enabled by short shaping blocklengths was intuitively attributed to the fact that the generated symbol sequences have fewer clusters of identical symbols \cite{fehenberger2019analysis,fehenberger2020impact}. However, a guiding principle for the temporal structure of symbol sequence was missing. Recently, we proposed in \cite{kaiquan2021EDI} a metric called energy dispersion index (EDI) to accurately indicate the NLI of correlated symbol sequences, where strong correlation between EDI and effective signal-to-noise ratio (SNR) was reported. The main contribution of the current paper is to propose the first NLI-tolerant signaling scheme designed based on the EDI.

In this paper, we first review PAS and CCDM as well as EDI in Sec.~\ref{sec:Pre}. Then, in Sec.~\ref{sec:LCCDM}, an amplitude shaper referred to as list-encoding CCDM (L-CCDM) is proposed. By exploiting the EDI diversity of the symbol sequences, L-CCDM improves NLI tolerance through EDI selection. In Sec.~\ref{sec:Sim}, the numerical results based on a multi-channel multi-span system show that with 256-ary quadrature amplitude modulation (256QAM), L-CCDM offers an effective SNR gain of $0.35$ dB with respect to CCDM. This SNR gain translates into an $8\%$ reach extension. Finally, conclusions are drawn in Sec.~\ref{sec:Conc}.

\section{Preliminaries}\label{sec:Pre}
\subsection{PAS and CCDM}

\begin{figure}[!t]
\centering
\includegraphics[width=\linewidth]{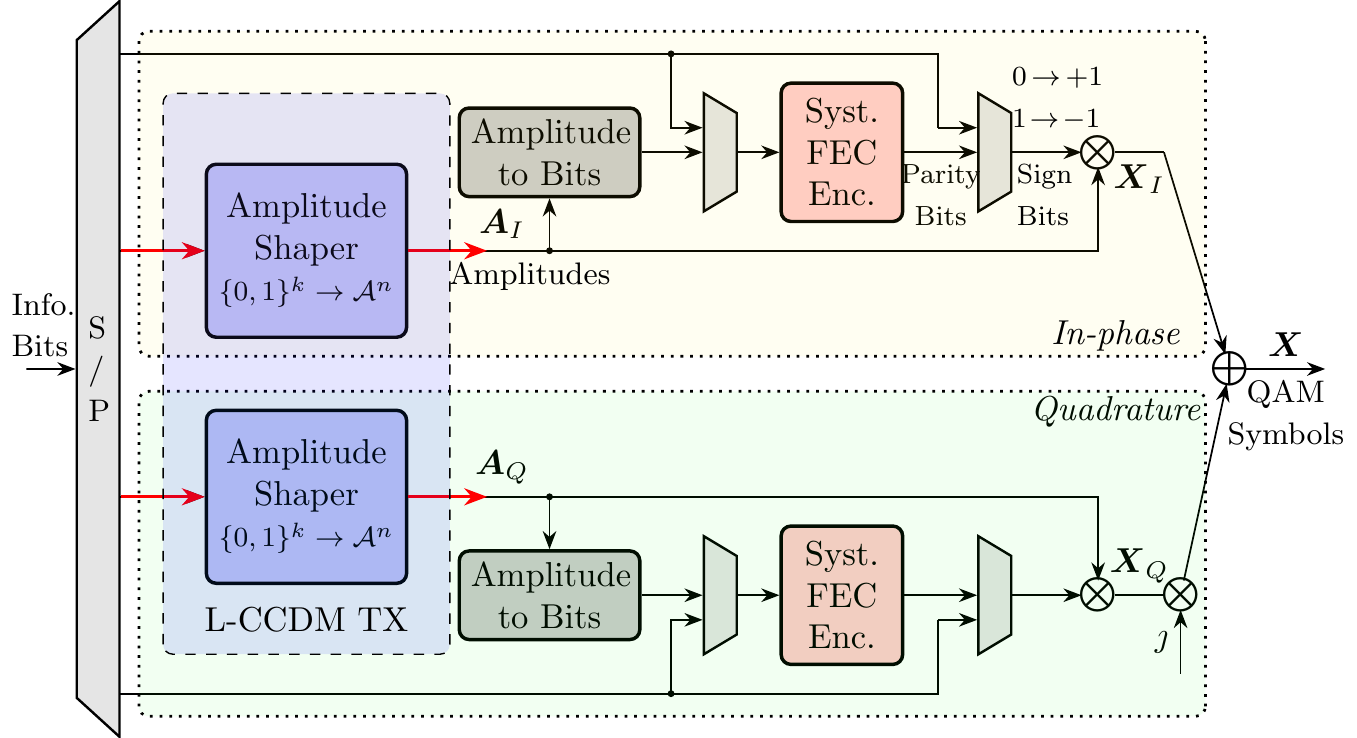}
\caption{Block diagram of PAS at the transmitter using 1D symbol mapping, where $\jmath$ is the imaginary unit. L-CCDM is to be placed at the dashed blue area.}
\label{blockListPos}
\end{figure}

Fig.~\ref{blockListPos} displays a standard PAS architecture. The top and bottom branches show the generation of shaped pulse amplitude modulation (PAM) symbols for in-phase and quadrature dimensions of QAM symbols, respectively. This PAS structure corresponds to the so-called ``1D symbol mapping strategy''~\cite[Fig. 3(a)]{skvortcov2020huffman}. For each branch, the generation of shaped PAM symbols, relying on an amplitude shaper and a systematic forward error correction (FEC) engine, can generally be described as follows. A sequence of information bits go through the amplitude shaper to yield nonuniformly distributed amplitudes. Next, the bits that represent these amplitudes, along with the top sequence of information bits, are fed into the FEC engine. The parity bits of the FEC codewords and the top sequence of information bits are transformed into sign bits, i.e., using $(0,1)\to(+1,-1)$. PAM symbols are generated by combining these signs with the amplitudes. Finally, two sequences of PAM symbols are assigned as the real and imaginary parts of a sequence of QAM symbols.

The amplitude sequences in the CCDM codebook $\mathcal{C}$ have a constant amplitude composition. Given an amplitude set $\mathcal{A}$ and blocklength $n$, the number of amplitudes $a$ for each codeword is a constant denoted by $n_{a}$, and thus, $n = \sum_{a\in\mathcal{A}} n_{a}$. For any CCDM codeword $\boldsymbol{A} = (A_1, A_2, \dotsc, A_n)$, the empirical amplitude distribution is fixed to be $\Prob_{A}(a)=n_{a}/n$ for $a\in\mathcal{A}$.

The shaping rate is defined as $R_s \triangleq k/n$ bits per amplitude (bit/amplitude). A finite-blocklength amplitude shaper will cause a non-zero rate loss. Let $H(\Prob_{A})$ denote the entropy of $\Prob_{A}$. Then the rate loss (in bits per amplitude) is defined in \cite[Eq.~(2)]{amari2019introducing} as 
\begin{equation}
\label{RateLoss}
  R_{L} \triangleq H(\Prob_{A})-\frac{k}{n}.
\end{equation}


\begin{figure}[!t]
    \centering
    \resizebox{1\linewidth}{!}{\tikzstyle{symBlk0} = [rectangle,draw=black, line width = 0.7pt, minimum height=30pt, minimum width=38pt,inner sep = 0pt]

\tikzstyle{symBlk} = [rectangle, minimum height=10pt, minimum width=10pt,inner sep = 0.3pt]

\begin{tikzpicture}

\foreach \i in {1,2,...,10}
\node[symBlk0] (sym\i) at (38*\i pt,0) {};

\node[symBlk] () at (sym1){$X_1$};
\node[symBlk] () at (sym2){$X_2$};
\node[symBlk] () at (sym3){$...$};
\node[symBlk] () at (sym4){$X_W$};
\node[symBlk] () at (sym5){$X_{W+1}$};
\node[symBlk] () at (sym6){$...$};
\node[symBlk] () at (sym7){$X_{n-W}$};
\node[symBlk] () at (sym8){$...$};
\node[symBlk] () at (sym9){$X_{n-1}$};
\node[symBlk] () at (sym10){$X_{n}$};

\draw [decorate, line width = 1 pt, decoration={brace,amplitude=10pt,raise=3pt}] (sym1.north west) -- (sym4.north east) node [midway,above=12pt] {$G_{1+W/2}^W=|X_1|^2+|X_2|^2+\ldots+|X_W|^2$};

\draw [decorate, line width = 1 pt, decoration={brace,amplitude=10pt,mirror,raise=3pt}] (sym2.south west) -- (sym5.south east) node [midway,below=12pt] { $G_{2+W/2}^W$};

\draw [decorate, line width = 1 pt, decoration={brace,amplitude=10pt,raise=3pt}] (sym7.north west) -- (sym10.north east) node [midway,above=12pt] {$G_{n-W/2}^W$};

\node[above=14pt] () at (sym5.north east){$......$};

\end{tikzpicture}}
    \caption{Window energy sequence $[G_{1+W/2}^W,G_{2+W/2}^W,\ldots,G_{n-W/2}^W]$ based on the symbol sequence $[X_1,X_2,\ldots,X_n]$.}
    \label{wndEngExmp}
\end{figure}

\begin{figure*}[t!]
\centering
\setkeys{Gin}{width=0.24\textwidth}
\subfloat[L-CCDM in the PAS transmitter.
\label{depExample_b}]{\includegraphics[width=0.45\linewidth]{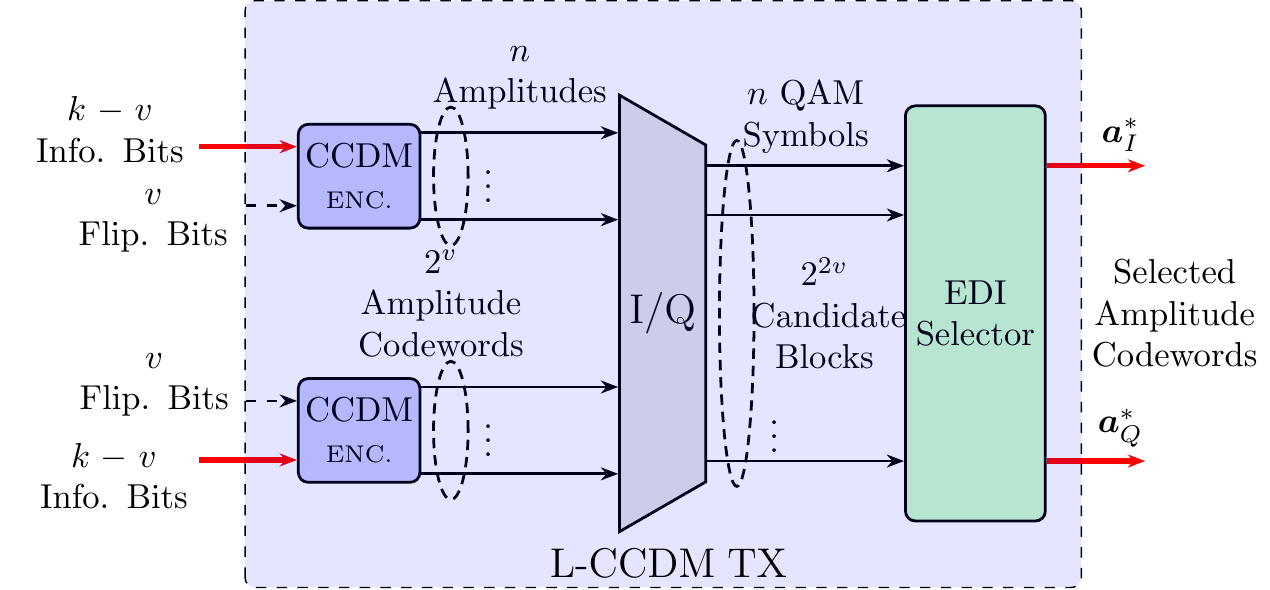}}
\hfill
\subfloat[L-CCDM in the PAS receiver.
\label{depExample_c}]{\includegraphics[width=0.26\linewidth]{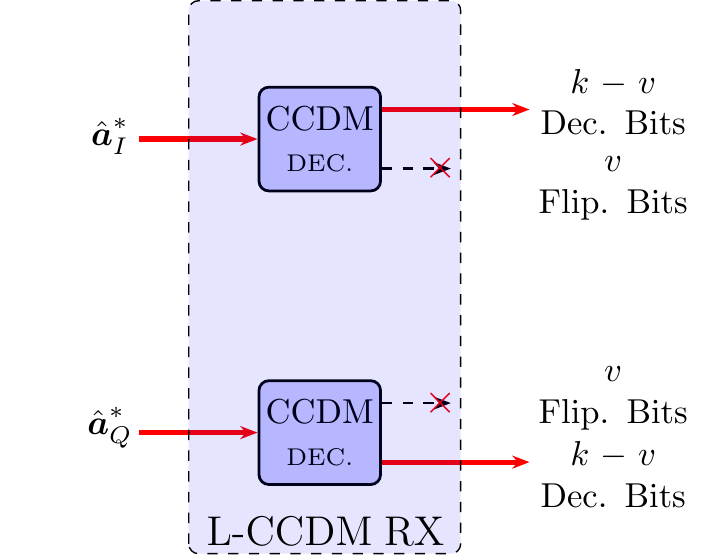}}
\hfill
\subfloat[EDI vs. flipping bit number $v$.
\label{depExample_c}]{\resizebox{0.28\linewidth}{!}{\definecolor{mycolor1}{rgb}{0.00000,0.44700,0.74100}%
\definecolor{mycolor2}{rgb}{0.85000,0.32500,0.09800}%
\begin{tikzpicture}

\begin{axis}[%
width=3.1647in,
height=2.4962in,
at={(0.758in,0.481in)},
scale only axis,
xmin=0,
xmax=4,
xlabel style={font=\color{white!15!black}},
xlabel={$v$},
xtick={0,1,2,3,4},
xticklabels={$0$,$1$,$2$,$3$,$4$},
ymin=-6.5,
ymax=-2.5,
ylabel style={font=\color{white!15!black}},
ylabel={EDI $\hat{\Psi}$ [dB]},
axis background/.style={fill=white},
xmajorgrids,
ymajorgrids,
legend style={legend cell align=left, align=left, draw=white!15!black}
]
\addplot [color=mycolor1, mark=*, mark options={solid, mycolor1, fill=white!60!mycolor1}, line width=1.5pt]
  table[row sep=crcr]{%
0	-2.71137686416995\\
1	-3.88196220644932\\
2	-4.80231166953095\\
3	-5.6328090906937\\
4	-6.22574830627187\\
};
\addlegendentry{Prefix Flipping Bits}

\addplot [color=mycolor2, mark=square*, mark options={solid, mycolor2, fill=white!60!mycolor2}, line width=1.5pt]
  table[row sep=crcr]{%
0	-2.71137686416995\\
1	-3.45140317093313\\
2	-3.8921105466292\\
3	-4.31585459385628\\
4	-4.57361850916861\\
};
\addlegendentry{Suffix Flipping Bits}

\end{axis}

\end{tikzpicture}%
}}
\caption{(a) Block diagram of L-CCDM in the PAS transmitter. Given $v$ flipping bits, each CCDM outputs $2^v$ possible amplitude codewords. Two sets of $2^v$ amplitudes codewords yield $2^{2v}$ candidate QAM symbol blocks. The corresponding optimal amplitude codewords $\boldsymbol{a}^*_{I}$ and $\boldsymbol{a}^*_{Q}$ are chosen for transmission. (b) Block diagram of the L-CCDM receiver. Given estimated amplitude codewords $\hat{\boldsymbol{a}}^*_{I}$ and $\hat{\boldsymbol{a}}^*_{Q}$, L-CCDM discard the decoded $v$ flipping bits and output $k-v$ decoded information bits. (c) EDI vs. flipping bit number $v$, when the flipping bits in L-CCDM ($n=180$, $R_s=1.85$, $W=10$) are appended as the prefix and the suffix of the binary inputs, respectively.}
\label{block}
\end{figure*}
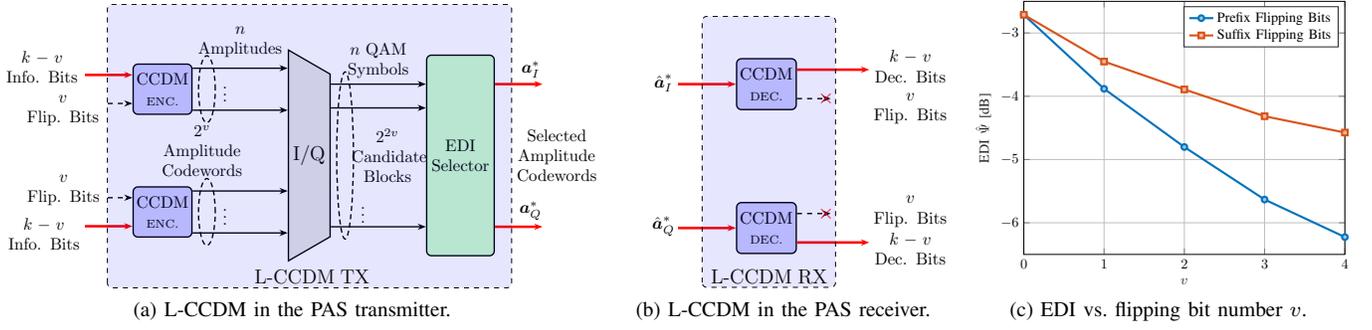

\subsection{Energy Dispersion Index}

EDI was designed to be a moving window statistic that measures the windowed energy fluctuation in a sequence of symbols $\boldsymbol{X}=[\ldots,X_i,X_{i+1},\ldots]$. For a given window length $W$ ($W$ is an even natural number), let $G_{i}^{W}$ denote the sum of $W+1$ symbol energies centered around the symbol $X_i$, i.e., $\WE_{i}\triangleq \sum_{j=i-W/2}^{i+W/2} |X_{j}|^2$. Based on this windowed energy process, EDI is defined as the ratio of the variance of the windowed energy to the mean of the windowed energy, both averaged over $i = 1, 2,\cdots, n$ \cite{kaiquan2021EDI}, i.e.,
\begin{equation}\label{EDI}
\Psi \triangleq\frac{\AvgVar{\WE}}{\AvgExp{\WE}},
\end{equation}
where 
\begin{align}
 \AvgExp{\WE} &\triangleq \frac{1}{n} \sum_{i=1}^{n}  \Exp{\WE_i}, \label{AvgExpWe}\\
 \AvgVar{\WE} &\triangleq  \frac{1}{n} \sum_{i=1}^{n} \Var{\WE_i}. \label{AvgVarWe}
\end{align}
The expressions in \eqref{AvgExpWe}--\eqref{AvgVarWe} show that EDI takes non-i.i.d. symbols into account.

In general, with a proper window length, the lower the EDI of transmitted symbols, the less NLI is to be induced \cite{kaiquan2021EDI}. 
EDI classifies phase-shift keying (PSK) as one of the ``best'' NLI-tolerant modulation formats, since the total energy within the window is always the same, yielding zero energy variations. For QAM modulation formats, if the transmitted symbols $\boldsymbol{X}$ are i.i.d., the EDI is simply determined by average symbol energy and kurtosis \cite[Eq.~(30)]{kaiquan2021EDI}. For the more general case of non-i.i.d. QAM symbols, EDI is also dependent on the window length $W$ since the correlations between symbols within the window can vary for different $W$. 

In the case of non-i.i.d. QAM symbols, an accurate indication of the NLI magnitude is obtained by EDI with an optimized window length $W$. However, \cite[Fig.~10]{kaiquan2021EDI} shows that the EDI is relatively insensitive to the $W$, which means that it can still show the general impact of the energy variations on the NLI without the optimization of $W$.

EDI in \eqref{EDI} describes the asymptotic energy variation behavior of the transmitted symbol sequences. In practice, we are more interested in the NLI-tolerance of one realization of symbol sequence, and thus the EDI of an individual sequence should be measured. Fig.~\ref{wndEngExmp} shows that given a realization of windowed energy sequence $\boldsymbol{G}=\boldsymbol{g}=[g_{1+W/2}^W,g_{2+W/2}^W,\ldots,g_{n-W/2}^W]$ is obtained based on a realization of an $n$-length symbol sequence $\boldsymbol{X}=\boldsymbol{x}=[x_1,x_2,\ldots,x_n]$. Note that the window length $W$ should be smaller than the blocklength $n$. Then, given $\boldsymbol{g}$, the EDI $\hat{\Psi}$ can be estimated numerically as
\begin{equation}
\label{EDIEst}
\hat{\Psi}(\boldsymbol{x}) \triangleq \frac{\TVar{\WE}}{\TExp{\WE}},
\end{equation} 
where $\TExp{\WE}$ is the estimated windowed energy mean, i.e.,
\begin{equation}
\label{WndMeanEst}
\TExp{\WE} = \frac{\sum_{i=1+W/2}^{n-W/2}g_{i}^W}{n-W},
\end{equation} 
and $\TVar{\WE}$ is the estimated windowed energy variance, i.e.,
\begin{equation}
\label{WndVarEst}
\TVar{\WE} = \frac{\sum_{i=1+W/2}^{n-W/2}(g_{i}-\TExp{\WE})^2}{n-W-1}.
\end{equation}




\section{List-Encoding CCDM}\label{sec:LCCDM}

In this section, we will introduce L-CCDM as a method to generate QAM symbol sequences with low EDI. The amplitude sequence codebook $\mathcal{C}$ of a standard CCDM with blocklength $n$ and shaping rate $R_s = k/n$ consists of $2^k$ amplitudes codewords. Even for small $n$, e.g., $n\geq 16$ and $R_s \geq 1.5$, there are usually $2^k \geq 2^8$ sequences with a wide variety of temporal structures, and thus, also potentially large EDI variations. Therefore, it is reasonable to assume that some of the codewords are more beneficial to NLI fiber transmission than the others. Based on this observation of EDI diversity in the codebook, a straightforward approach to realize NLI mitigation is to select and use a subset of NLI-tolerant (low-EDI) symbol sequences.  

\subsection{Principle}

In the PAS architecture using 1D symbol mapping as shown in Fig.~\ref{blockListPos}, the proposed L-CCDM replaces the amplitude shapers with red inputs and outputs. 
Fig.~\ref{block}(a) displays the block diagram of L-CCDM at the transmitter. L-CCDM is built based on standard CCDMs, and it selects amplitude codewords that generate low-EDI symbol sequences.
For any given $k-v$ bit input sequence of a CCDM, we insert $v$ flipping bits.
As flipping bit changes between 0 and 1, CCDM produces $2^v$ candidate amplitude sequences from its codebook depending on the values of these flipping bits. Therefore, the shaping rate of L-CCDM is thus $R_s=(k-v)/n$.
We note that this can be done by using a single CCDM $2^v$ times, or using $2^v$ CCDMs at once.
We denote these candidate sequences at the outputs of CCDMs in Fig.~\ref{blockListPos} by $\boldsymbol{a}^{i}_{I}=[a^{i}_{I,1},a^{i}_{I,2},\ldots,a^{i}_{I,n}]$ and $\boldsymbol{a}^{j}_{Q}=[a^{j}_{Q,1},a^{j}_{Q,2},\ldots,a^{j}_{Q,n}]$ for the in-phase and quadrature branches, respectively, where $i, j = 1, 2,\dotsc, 2^v$. 

Then L-CCDM generates $2^{2v}$ QAM symbol sequence candidates considering all possible combinations of $\boldsymbol{a}^{i}_{I}$ and $\boldsymbol{a}^{j}_{Q}$.
Since these candidates will only be used to compute EDIs, this generation requires no sign bits from the systematic FEC engine.
As a result, these ``pseudo'' QAM symbol sequence candidates are given as $\boldsymbol{x}_{i,j}=\boldsymbol{a}_{I}^{i}+\jmath\boldsymbol{a}_{Q}^{j}$. 
We denote the set of $2^{2v}$ QAM symbol sequence candidates by $\mathcal{S}=\{\boldsymbol{x}_{i,j}: (i,j)\in\{1,2,\ldots,2^v\}^2\}$.
Next, the EDIs of all the candidate symbol blocks are measured, and L-CCDM chooses the ``best'' candidate $\boldsymbol{x}^{*}$ that has the smallest EDI, i.e., 

\begin{equation}
    \boldsymbol{x}^{*} = \argmin_{\boldsymbol{x}\in \mathcal{S}} \hat{\Psi}(\boldsymbol{x}).
\end{equation}

Finally, the selected QAM symbol block $\boldsymbol{x}^{*}=\boldsymbol{a}^{*}_{I}+\jmath\boldsymbol{a}^{*}_{Q}$ determines the output amplitude codewords $\boldsymbol{A}_{I}=\boldsymbol{a}^{*}_{I}$ and $\boldsymbol{A}_{Q}=\boldsymbol{a}^{*}_{Q}$, which are later processed by standard PAS procedure as explained in Fig.~\ref{blockListPos}. Fig.~\ref{block}(b) shows that at the receiver, the only modification required is to merely discard the flipping bits after the standard CCDM decoding of estimated amplitude codewords $\hat{\boldsymbol{a}}^*_{I}$ and $\hat{\boldsymbol{a}}^*_{Q}$. With minor changes on the PAS architecture, L-CCDM generates symbol blocks with lower EDI than that of the corresponding CCDM. 

L-CCDM is designed to only improve the blockwise EDI, i.e., symbols from previous and future codewords are not taken into account when computing the EDI. One drawback of L-CCDM is therefore that \emph{inter-block effects} are neglected. However, such inter-block effects are expected to be marginal if the blocklength is relatively large. 

\subsection{Flipping Bits Insertion and Rate Loss}
The QAM symbol sequence candidates are desired to be as different as possible in terms of their EDI, in order to maximize the gains in NLI-tolerance that we can obtain. 
A CCDM indexes constant composition sequences using a lexicographical ordering for both its input and output \cite{schulte2015constant}.
As a result, compared to the bits at the end of the binary input, flipping the bits at the beginning is more important in changing the temporal structure of two candidate amplitude codewords. 
In the next example, we will show that appending the flipping bits to the binary inputs as a prefix is more efficient to reduce EDI using L-CCDM.

\begin{table}[]
\renewcommand{\arraystretch}{1.3}
\caption{A Mapping Example of CCDM's Input and Output.}
\label{ExmpTab2}
\centering
\begin{tabular}{ |c|c|  }
 \hline
 Bits Input & Amplitudes Output \\ 
 \hline
  $1111111111$ & $7553331111$  \\
  $\vdots$ &  $\vdots$  \\
  $1000000000$ & $3153153117$  \\
  $\vdots$ &  $\vdots$  \\
  $0000000001$ & $1111333575$  \\
  $0000000000$ & $1111333557$  \\
\hline
\end{tabular}
\end{table}

\textit{Example:} Given a CCDM with $k=10$, $n=10$, and amplitude composition $[4, 3, 2, 1]$ for the amplitudes $\{1,3,5,7\}$, the mapping between the binary inputs and amplitude codewords is shown in Table \ref{ExmpTab2}. For $v=1$ flipping bit, if $k-v=9$ information bits are all zeros, the two amplitude codewords corresponding to bit sequences $[0,\ldots,0,B]$, $B\in\{0,1\}$ only differ two amplitudes at the end. By comparison, the two amplitude codewords corresponding to $[B,0,\ldots,0]$ have more difference. Fig.~\ref{block}(c) also clearly shows that when $n=180$, with respect to the suffix approach, the prefix approach reduces the EDI as $v$ increases. For $v=4$ flipping bits, the average EDI obtained with the prefix approach is $1.65$ dB less that the suffix approach.

The inserted flipping bits cause an additional rate loss with respect to standard CCDM. The rate loss of L-CCDM is
\begin{equation} \label{RateLoss2}
   R_{L}^{\textrm{List}}  = \underbrace{H(\Prob_{A})- \frac{k}{n}}_{\text{Eq.~\eqref{RateLoss}}} 
   + \underbrace{\frac{v}{n}}_{\text{Flipping bit loss}},
\end{equation}
which includes an extra rate loss term $v/n$ that accounts for the effect of flipping bits. Note that $R_{L}$ in \eqref{RateLoss} can be seen as a special case of $R_{L}^{\textrm{List}}$ when $v=0$.

In L-CCDM, $v$ and $n$ can be used to trade performance for complexity or vice versa. 
For example, for a fixed $v$, if $n$ is large enough, the additional rate loss $v/n$ of L-CCDM in \eqref{RateLoss2} is negligible. 
However, increasing $n$ typically causes an increased complexity in shaping and deshaping~\cite{gultekin2019probabilistic}. 
On the other hand, for a fixed $n$, using more flipping bits means more QAM symbol sequence candidates to select among and a possibly larger reduction in average EDI. 
This increase in $v$, however, requires extra $2^v-1$ CCDM encoding instances. 
The EDI measurement also requires some computational power. 
Therefore, the parameters of L-CCDM should be chosen carefully in order to reach a favorable trade-off between performance and complexity. 


\begin{table}[]
    \centering
    \caption{Simulation Parameters.}
    \label{FiberParam}
    \begin{tabular}{c|c} 
    \hline\hline\textbf{Parameter} & \textbf{Value} \\
    \hline Modulation & $256$QAM \\
    Polarization & Single \\
    Center wavelength ($\lambda$) &  1550 nm \\
    Symbol rate & $32$ GBd \\
    WDM spacing ($\Delta f$) & $50$ GHz \\
    \# WDM channels ($N_{\mathrm{ch}}$) & 11 \\
    Pulse shape & Root-raised cosine \\
    Pulse roll-off & $10\% $ \\
    \hline Span length & $80$ km\\
    Fiber loss & $0.2$ dB/km \\
    Dispersion parameter ($D$) & $17$ ps/nm/km \\
    Nonlinear parameter ($\gamma$) & $1.37$ 1/W/km \\
    EDFA noise figure & $6$ dB \\
    \hline Oversampling factor & $2$ \\\hline\hline
    \end{tabular}
\end{table}

\begin{table}[]
    \centering
    \caption{Parameters of 256QAM Uniform Signaling and PAS.}
    \label{PASParam}
    \begin{tabular}{c|c|c|c|c|c|c}
    \hline\hline\textbf{Parameter} & \multicolumn{2}{c|}{\textbf{Uniform}} & \multicolumn{4}{c}{\textbf{PAS}} \\\hline
    $R_c$ & $2/3$ & $3/5$ & \multicolumn{4}{c}{$4/5$} \\\hline
    $R_s$ & \multicolumn{2}{c|}{-} & 2.2 & 2.3 & 2.4 & 2.5 \\\hline
    Total $R$ (bit/4D-sym) & $9.6$ & $10.7$ & $9.6$ & $10$ & $10.4$ & $10.8$ \\\hline
    Info. Rate (Gbps) & $307$ & $342$ & $307$ & $320$ & $333$ & $346$ \\\hline\hline
    \end{tabular}
\end{table}

\begin{figure*}[]
\centering
{\resizebox{0.48\linewidth}{!}{\definecolor{mycolor1}{rgb}{0.00000,0.44700,0.74100}%
\definecolor{mycolor2}{rgb}{0.85000,0.32500,0.09800}%
\definecolor{mycolor3}{rgb}{0.92900,0.69400,0.12500}%
\definecolor{mycolor4}{rgb}{0.49400,0.18400,0.55600}%
\definecolor{mycolor5}{rgb}{0.46600,0.67400,0.18800}%
\begin{tikzpicture}[font=\large]

\begin{axis}[name=mainplot,
width=4.521in,
height=3.566in,
at={(0.758in,0.481in)},
scale only axis,
xmin=-6.5,
xmax=-1,
xlabel style={font=\large\color{white!15!black}},
xlabel={Launch Power [dBm]},
ymin=14.5,
ymax=17,
ylabel style={font=\large\color{white!15!black}},
ylabel={Effective SNR [dB]},
axis background/.style={fill=white},
title style={font=\bfseries},
xmajorgrids,
ymajorgrids,
legend style={at={(0.01,0.005)}, anchor=south west, legend cell align=left, align=left, draw=white!15!black}
]
\addplot [color=mycolor1, line width=1pt, mark=star, mark options={solid, fill=white!60!mycolor1}
]
  table[row sep=crcr]{%
-6.5	15.1445984247618\\
-6	15.5197207372625\\
-5.5	15.8407068984313\\
-5	16.1026060910459\\
-4.5	16.3044463183528\\
-4	16.421524202965\\
-3.5	16.4168816717157\\
-3	16.300397475549\\
-2.5	16.0467005460235\\
-2	15.7231847905006\\
-1.5	15.1850070109583\\
-1	14.6061966535861\\
};
\addlegendentry{$v=0$}

\addplot [color=mycolor2, line width=1pt, mark=pentagon*, mark options={solid, fill=white!60!mycolor2}
]
  table[row sep=crcr]{%
-6.5	15.1713617371801\\
-6	15.5551325299959\\
-5.5	15.8880857525484\\
-5	16.1695194150833\\
-4.5	16.3905517436819\\
-4	16.5265027578309\\
-3.5	16.5736754905179\\
-3	16.4802024126838\\
-2.5	16.2836933338119\\
-2	15.9423919190426\\
-1.5	15.4834208645702\\
-1	14.9042138052923\\
};
\addlegendentry{$v=1$}

\addplot [color=mycolor3, line width=1pt, mark=square*, mark options={solid, fill=white!60!mycolor3}
]
  table[row sep=crcr]{%
-6.5	15.1950370458433\\
-6	15.580631061916\\
-5.5	15.9311930058069\\
-5	16.2243520164232\\
-4.5	16.4645456104302\\
-4	16.6148549980152\\
-3.5	16.6755958456197\\
-3	16.6050146857091\\
-2.5	16.4236548912407\\
-2	16.1210935590108\\
-1.5	15.6915074473434\\
-1	15.108907825497\\
};
\addlegendentry{$v=2$}

\addplot [color=mycolor4, line width=1pt, mark=triangle*, mark options={solid, fill=white!60!mycolor4}
]
  table[row sep=crcr]{%
-6.5	15.2061104599356\\
-6	15.5986421784612\\
-5.5	15.9490581947594\\
-5	16.2627751661756\\
-4.5	16.4994904040038\\
-4	16.6696456771297\\
-3.5	16.7308754060521\\
-3	16.6929052654993\\
-2.5	16.5360413742094\\
-2	16.2279923407094\\
-1.5	15.8046640573045\\
-1	15.2650327082356\\
};
\addlegendentry{$v=3$}

\addplot [color=mycolor5, line width=1pt, mark=*, mark options={solid, fill=white!60!mycolor5}
]
  table[row sep=crcr]{%
-6.5	15.2162846499886\\
-6	15.6095611454737\\
-5.5	15.9687985263163\\
-5	16.271821365704\\
-4.5	16.5299583670542\\
-4	16.7049754516068\\
-3.5	16.7722216079052\\
-3	16.7448841599882\\
-2.5	16.5878446966044\\
-2	16.3106908599726\\
-1.5	15.9224879371761\\
-1	15.3727926469235\\
};
\addlegendentry{$v=4$}

\addplot [color=mycolor1,line width=2.5pt,mark=triangle, mark options={solid, rotate=180, fill=mycolor1}]
  table[row sep=crcr]{%
  -4	16.4168816717157\\
};
\addplot [color=mycolor5,line width=2.5pt,mark=triangle, mark options={solid, fill=mycolor5}]
  table[row sep=crcr]{%
  -3.5	16.7722216079052\\
 };

\draw[-, very thick, dashed, draw opacity=0.9] (-4,16.421524202965) -- (-5,16.421524202965) node[] {};

\draw[-, very thick, dashed, draw opacity=0.9] (-3.5,16.7722216079052) -- (-5,	16.7722216079052) node[] {};

\draw[<->, very thick, draw opacity=0.9] (-5,16.7722216079052) -- (-5,	16.421524202965) node[xshift=-19pt, yshift=17pt, text width = 1.5cm] {$0.35$ dB};

\end{axis}
\node[font=\Large] at (13,10) {(a)};

\begin{axis}[%
at={(mainplot.south east)},
anchor=south east,
yshift=1cm,
xshift=-3cm,
width=0.26\textwidth,
scale only axis,
xmin=0,
xmax=4,
xlabel style={at={(axis description cs:0.5,-0.1)},anchor=north, font=\normalsize	\color{white!15!black}},
xlabel={$v$},
xtick distance=1,
ymin=-5,
ymax=-0,
ylabel style={at={(axis description cs:-0.24,0.45)},anchor=north, font=\normalsize	\color{white!15!black}},
ylabel={EDI $\hat{\Psi}$ [dB]},
xmajorgrids,
ymajorgrids,
axis background/.style={fill=white},
legend style={legend cell align=left, align=left, draw=white!15!black}
]
\addplot [color=mycolor1, mark=*, mark options={solid, mycolor1, fill=white!60!mycolor1}, line width=1.5pt]
  table[row sep=crcr]{%
0	-0.658001778671189\\
1	-1.75365096048217\\
2	-2.73802866089263\\
3	-3.43746264593544\\
4	-4.034756887524\\
};

\node[] at (0.7,-0.66) {$-0.66$ dB};
\node[] at (3.2,-4.1) {$-4.03$ dB};

\end{axis}

\end{tikzpicture}}}
\hfil
{\resizebox{0.48\linewidth}{!}{\input{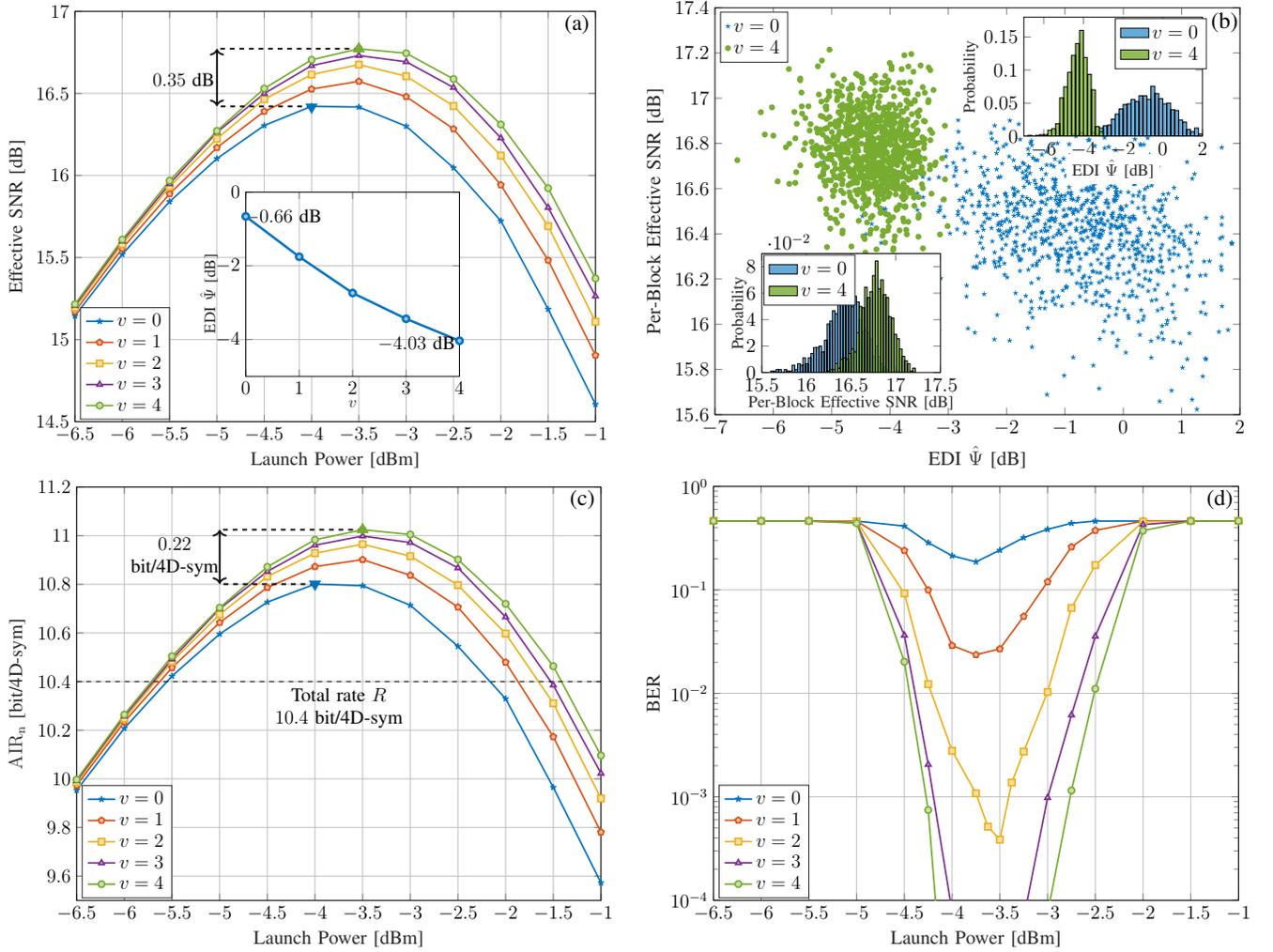}}}
{\resizebox{0.48\linewidth}{!}{\definecolor{mycolor1}{rgb}{0.00000,0.44700,0.74100}%
\definecolor{mycolor2}{rgb}{0.85000,0.32500,0.09800}%
\definecolor{mycolor3}{rgb}{0.92900,0.69400,0.12500}%
\definecolor{mycolor4}{rgb}{0.49400,0.18400,0.55600}%
\definecolor{mycolor5}{rgb}{0.46600,0.67400,0.18800}%
\begin{tikzpicture}[font=\large]

\begin{axis}[name=mainplot,
width=4.521in,
height=3.566in,
at={(0.758in,0.481in)},
scale only axis,
xmin=-6.5,
xmax=-1,
xlabel style={font=\large\color{white!15!black}},
xlabel={Launch Power [dBm]},
ymin=9.5,
ymax=11.2,
ylabel style={font=\large\color{white!15!black}},
ylabel={$\mathrm{AIR}_{\mathrm{n}}$ [bit/4D-sym]},
axis background/.style={fill=white},
title style={font=\bfseries},
xmajorgrids,
ymajorgrids,
legend style={at={(0.01,0.005)}, anchor=south west, legend cell align=left, align=left, draw=white!15!black}
]
\addplot [color=mycolor1, line width=1pt, mark=star, mark options={solid, fill=white!60!mycolor1}
]
  table[row sep=crcr]{%
-6.5	9.95283128781238\\
-6	10.2076670724132\\
-5.5	10.4224995997901\\
-5	10.5954364246123\\
-4.5	10.72681505671\\
-4	10.8010371251889\\
-3.5	10.7943659428964\\
-3	10.71395581475\\
-2.5	10.5448695512306\\
-2	10.3294909946416\\
-1.5	9.96486492402685\\
-1	9.5724380407546\\
};
\addlegendentry{$v=0$}

\addplot [color=mycolor2, line width=1pt, mark=pentagon*, mark options={solid, fill=white!60!mycolor2}
]
  table[row sep=crcr]{%
-6.5	9.97287421389854\\
-6	10.2338636958886\\
-5.5	10.4569276047093\\
-5	10.6426986107327\\
-4.5	10.7868134571179\\
-4	10.8730333554568\\
-3.5	10.9017291284222\\
-3	10.837110018443\\
-2.5	10.7063499621957\\
-2	10.4793646606038\\
-1.5	10.1729814994759\\
-1	9.78050502141693\\
};
\addlegendentry{$v=1$}

\addplot [color=mycolor3, line width=1pt, mark=square*, mark options={solid, fill=white!60!mycolor3}
]
  table[row sep=crcr]{%
-6.5	9.98597264969602\\
-6	10.2475823952514\\
-5.5	10.4823225865216\\
-5	10.6760494022708\\
-4.5	10.8323615244899\\
-4	10.9282480232254\\
-3.5	10.9649426927446\\
-3	10.915959949931\\
-2.5	10.7968253624177\\
-2	10.5973521999738\\
-1.5	10.3108613886221\\
-1	9.91923692252063\\
};
\addlegendentry{$v=2$}

\addplot [color=mycolor4, line width=1pt, mark=triangle*, mark options={solid, fill=white!60!mycolor4}
]
  table[row sep=crcr]{%
-6.5	9.99126285600122\\
-6	10.2575285656517\\
-5.5	10.492229768208\\
-5	10.6992993723708\\
-4.5	10.8529088492175\\
-4	10.9611342915782\\
-3.5	10.9987707793663\\
-3	10.9719671423968\\
-2.5	10.8676843394349\\
-2	10.6656097086558\\
-1.5	10.3853643067947\\
-1	10.0226021916382\\
};
\addlegendentry{$v=3$}

\addplot [color=mycolor5, line width=1pt, mark=*, mark options={solid, fill=white!60!mycolor5}
]
  table[row sep=crcr]{%
-6.5	9.9967997483853\\
-6	10.264089211689\\
-5.5	10.5043444430416\\
-5	10.7038352441102\\
-4.5	10.8719377347761\\
-4	10.9834409868507\\
-3.5	11.0247809415053\\
-3	11.0050618597016\\
-2.5	10.9020150070234\\
-2	10.7197832911908\\
-1.5	10.4633385203269\\
-1	10.0956319362315\\
};
\addlegendentry{$v=4$}

\draw[-, very thick, dashed, draw opacity=0.9] (-4,10.8010371251889) -- (-5,10.8010371251889) node[] {};

\draw[-, very thick, dashed, draw opacity=0.9] (-3.5,11.0247809415053) -- (-5,	11.0247809415053) node[] {};

\draw[<->, very thick, draw opacity=0.9] (-5,11.0247809415053) -- (-5, 10.8010371251889) node[xshift=-28pt, yshift=17pt, text width = 2cm,align=center] {$0.22$ bit/4D-sym};
\draw[-,very thick, dashed, draw opacity=0.5] (-6.5,10.4) -- (-1,10.4) node[align=center,midway,below,text width=3.6cm] {Total rate $R$\\ $10.4$ bit/4D-sym};

\addplot [color=mycolor1,line width=2.5pt,mark=triangle, mark options={solid, rotate=180, fill=mycolor1}]
  table[row sep=crcr]{%
  -4	10.8010371251889\\
};
\addplot [color=mycolor5,line width=2.5pt,mark=triangle, mark options={solid, fill=mycolor5}]
  table[row sep=crcr]{%
  -3.5	11.0247809415053\\
 };

\end{axis}
\node[font=\Large] at (13,10) {(c)};
\end{tikzpicture}%
}}
\hfil
{\resizebox{0.48\linewidth}{!}{\definecolor{mycolor1}{rgb}{0.00000,0.44700,0.74100}%
\definecolor{mycolor2}{rgb}{0.85000,0.32500,0.09800}%
\definecolor{mycolor3}{rgb}{0.92900,0.69400,0.12500}%
\definecolor{mycolor4}{rgb}{0.49400,0.18400,0.55600}%
\definecolor{mycolor5}{rgb}{0.46600,0.67400,0.18800}%
\begin{tikzpicture}[font=\large]

\begin{axis}[name=mainplot,
width=4.521in,
height=3.566in,
at={(0.758in,0.481in)},
scale only axis,
xmin=-6.5,
xmax=-1,
xlabel style={font=\large\color{white!15!black}},
xlabel={Launch Power [dBm]},
ymode=log,
ymin=0.0001,
ymax=1,
ylabel style={font=\large\color{white!15!black}},
ylabel={BER},
axis background/.style={fill=white},
title style={font=\bfseries},
xmajorgrids,
ymajorgrids,
legend style={at={(0.01,0.005)}, anchor=south west, legend cell align=left, align=left, draw=white!15!black}
]
\addplot [color=mycolor1,line width =1pt,mark=star, mark options={solid, fill=white!60!mycolor1}
]
  table[row sep=crcr]{%
-6.5	0.463573717948718\\
-6	0.463339031339031\\
-5.5	0.462726970560304\\
-5	0.462437559354226\\
-4.5	0.413905745489079\\
-4.25	0.285972198480532\\
-4	0.213907051282051\\
-3.75	0.18658896011396\\
-3.5	0.242003727445394\\
-3.25	0.319996082621083\\
-3	0.385170156695157\\
-2.75	0.440619112060779\\
-2.5	0.462558404558405\\
-2	0.462860873694207\\
-1.5	0.463278727445394\\
-1	0.464087606837607\\
};
\addlegendentry{$v=0$}

\addplot [color=mycolor2,line width =1pt,mark=pentagon*, mark options={solid, fill=white!60!mycolor2}
]
  table[row sep=crcr]{%
-6.5	0.464159483270405\\
-6	0.463548479019734\\
-5.5	0.463207356747643\\
-5	0.462501365438959\\
-4.5	0.240006411626416\\
-4.25	0.0996982261166916\\
-4	0.0289723112725891\\
-3.75	0.0235970411531429\\
-3.5	0.0268897200256465\\
-3.25	0.0554080644012253\\
-3	0.119715893709482\\
-2.75	0.259963524969723\\
-2.5	0.375512455177982\\
-2	0.462682909453587\\
-1.5	0.463480325805609\\
-1	0.463840208021657\\
};
\addlegendentry{$v=1$}

\addplot [unbounded coords=jump,color=mycolor3, line width =1pt,mark=square*, mark options={solid, fill=white!60!mycolor3}
]
  table[row sep=crcr]{%
-6.5	0.464329485535129\\
-6	0.463382143366111\\
-5.5	0.463290580019951\\
-5	0.459866514654886\\
-4.5	0.0925163966874101\\
-4.25	0.0122784072015581\\
-4	0.00278671638718667\\
-3.75	0.00107984577138062\\
-3.625 5.147641442211771e-04\\ 
-3.5	0.000386046585277026\\
-3.375 0.001373585577882\\ 
-3.25	0.002738888888889\\
-3	0.0103158638544487\\
-2.75	0.0668928039839753\\
-2.5	0.17366182129115\\
-2	0.45592418412427\\
-1.5	0.463337608664671\\
-1	0.463647095149874\\
};
\addlegendentry{$v=2$}

\addplot [unbounded coords=jump,color=mycolor4, line width =1pt,mark=triangle*, mark options={solid, fill=white!60!mycolor4}
]
  table[row sep=crcr]{%
-6.5	0.46398332264272\\
-6	0.463547264390754\\
-5.5	0.462967357993016\\
-5	0.445955265721141\\
-4.5	0.0363197166591436\\
-4.25	0.00204994100365065\\
-4	7.75940576026101e-05\\
-3.75	nan\\
-3.5	nan\\
-3.25	6.67094291212316e-05\\
-3	0.000981220155370252\\
-2.75	0.00616844447612034\\
-2.5	0.0355530610790393\\
-2	0.427895968450811\\
-1.5	0.463143515548904\\
-1	0.463450454945003\\
};
\addlegendentry{$v=3$}

\addplot [unbounded coords=jump,color=mycolor5,line width =1pt,mark=*, mark options={solid, fill=white!60!mycolor5}
]
  table[row sep=crcr]{%
-6.5	0.464160963786712\\
-6	0.463188385134493\\
-5.5	0.463035239045718\\
-5	0.441172298260622\\
-4.5	0.0201975453854196\\
-4.25	0.000745714919367614\\
-4	1e-6\\
-3.75	nan\\
-3.5	nan\\
-3.25	nan\\
-3	7.4505750403954e-05\\
-2.75	0.00115153423312106\\
-2.5	0.0110646564014827\\
-2	0.373930947628553\\
-1.5	0.463088465925292\\
-1	0.463576775021386\\
};
\addlegendentry{$v=4$}

\end{axis}
\node[font=\Large] at (13,10) {(d)};
\end{tikzpicture}%
}}
\caption{Performance analysis of L-CCDM and standard CCDM ($n=1800$) at $1600$ km. The 256QAM symbol sequences are generated by PAS using standard CCDM ($v=0$) or L-CCDM ($v=1,2,\ldots,4$ with window length $W=100$). (a) Effective SNR vs. launch power, where the inset shows EDI $\hat{\Psi}$ vs. $v$. (b) For $v=0$ and $v=4$, the scatter of per-shaping block effective SNR vs. EDI, and the insets further show the histogram of per-shaping block effective SNR and EDI. (c) AIR vs. launch power. (d) BER vs. launch power. The filled triangles in (a), (c) for $v=0,4$ correspond to the same triangle points as shown in Fig.~\ref{fig_blkLen}(a)--(b).}
\label{figMtx}
\end{figure*}

\section{Numerical Analysis}\label{sec:Sim}

\subsection{Simulation Setup}

The performance of PAS using 256QAM with standard CCDM or L-CCDM is analyzed in a multi-span wavelength division multiplexing (WDM) optical communication system. The key parameters of the fiber simulation are displayed in Table \ref{FiberParam}. The simulation is implemented by using the split-step Fourier method. The channel of interest is situated in the center of the WDM band. All channels use the same signaling scheme, and the signal on each channel is independently generated with the root-raised cosine pulse shaping. After propagation over each span of standard single-mode fiber, the attenuation is compensated by an Erbium-doped fiber amplifier (EDFA). At the receiver, the channel of interest is filtered with a matched filter, and chromatic dispersion compensation. 

The binary low-density parity check codes from the DVB-S2 standard~\cite{european2009digital} with codeword length $64800$ is used as the FEC code.
At the receiver, 50 decoding iterations are realized. The number of flipping bits $v=0$ corresponds to standard CCDM. Shaping blocklengths from $n=180$ to $n=5400$ are evaluated. We choose window length $W=100$ and note that the EDI measurement is insensitive to the value of $W$. 
At most $v=4$ flipping bits are considered, which generates up to $16$ amplitude sequence candidates and cause an extra rate loss of $2.22\times10^{-3}$ bit/amplitude, and thus, the complexity increase is not too much. 
To make a fair comparison between different values of $v$ \cite{vassilieva2020fairness}, Maxwell-Boltzmann distribution $\Prob_{A}$ with slightly larger entropy $H(\Prob_{A})$ is used by L-CCDM to cover the additional rate loss due to flipping bits $v/n$, such that L-CCDM has the same shaping rate $R_s=(k-v)/n$ as that $R_s=k/n$ of standard CCDM, see \eqref{RateLoss} and \eqref{RateLoss2}. 
For example, at $n=1800$, to achieve $R_s= 2.4$ bit/amplitude, CCDM and L-CCDM ($v=4$) have $H(\Prob_{A})=2.4189$ and $2.4205$ bits, respectively. Table~\ref{PASParam} shows the parameters for uniform signaling and PAS that achieve various rates.

To quantify the performance improvement obtained with L-CCDM, we evaluated the effective SNR, achievable information rate (AIR) and end-to-end bit-error rate (BER). 
Regarding AIR, we used ``finite-blocklength" bit-metric decoding (BMD) rate (bit/4D-sym) as the figure of merit for 256QAM PAS, which accounts for the rate loss $R^{\textrm{List}}_{L}$ in \eqref{RateLoss2} as ~\cite[Eq.~(8)]{amari2019introducing}) 
\begin{equation} \label{AIRN}
   \mathrm{AIR}_{\mathrm{n}}=4\left[H(X)-\sum_{i=1}^{4} H\left(B_{i} \mid Y\right)\right]-  4R^{\textrm{List}}_{L},
\end{equation}
where $B_{i}$ is a random variable representing the $i$-th binary label of the 16PAM symbol, and $Y$ is the received symbol.

\subsection{Performance at Fixed Transmission Distance}

Fig.~\ref{figMtx}(a) displays effective SNR vs. launch power at the transmission distance of $1600$ km for $n=1800$.
The total transmission rate is fixed at $R=10.4$ bit/4D-symbol with $R_s= 2.4$ bit/amplitude. 
The inset figure shows that the EDI of transmitted symbols indeed decreases when more flipping bits are used. While the EDI of standard CCDM ($v=0$) is $-0.66$~dB, L-CCDM suppresses the EDI down to $-4.03$~dB at $v=4$. 
This decrease in EDI results in an effective SNR gain of $0.35$~dB. The optimal launch power for L-CCDM with $v=4$ is $-3.5$~dBm, which is $0.5$~dB higher than that of CCDM. 

Fig.~\ref{figMtx}(b) further visualizes how the effective SNR is improved by L-CCDM.
Compared with CCDM, L-CCDM with $v=4$ on average pick $1/16$ NLI-tolerant amplitude sequences from codebook. To keep the same shaping rate, L-CCDM has to use a composition with slightly larger entropy than that of CCDM, and thus a larger amplitude sequence codebook. This allows L-CCDM to move the EDI distribution to a lower level, as shown in the top-right histogram inset.
As a result, the per-block effective SNR cloud of L-CCDM is located around a higher SNR value and is more concentrated, which can also be observed from the bottom left histograms inset. 

In Fig.~\ref{figMtx}(c), it can be seen that the AIR curves almost follow the same behaviour as the SNR curves in (a). This indicates that most of the effective SNR gains by L-CCDM translate into AIR gains, and the flipping rate loss is negligible. With $v=4$, the AIR gains can be up to $0.22$ bit/4D-symbol. 
Fig.~\ref{figMtx}(d) shows that thanks to the NLI reduction, L-CCDM outperforms standard CCDM in terms of BER. From $-4.25$ dBm to $-3$ dBm, L-CCDM with $v=3,4$ achieves BER below $1\times10^{-4}$.


\begin{figure*}[t!]
\centering
\setkeys{Gin}{width=0.24\textwidth}
\resizebox{0.48\linewidth}{!}{\definecolor{mycolor1}{rgb}{0.00000,0.44700,0.74100}%
\definecolor{mycolor2}{rgb}{0.85000,0.32500,0.09800}%
\definecolor{mycolor3}{rgb}{0.92900,0.69400,0.12500}%
\definecolor{mycolor4}{rgb}{0.49400,0.18400,0.55600}%
\definecolor{mycolor5}{rgb}{0.46600,0.67400,0.18800}%
\begin{tikzpicture}

\begin{axis}[%
width=4.521in,
height=3.566in,
at={(0.758in,0.481in)},
scale only axis,
xmode=log,
log ticks with fixed point,
xmin=100,
xmax=10000,
xminorticks=true,
xlabel style={font=\color{white!15!black}},
xlabel={Blocklength $n$},
ymin=15.95,
ymax=17.05,
ylabel style={font=\color{white!15!black}},
ylabel={Effective SNR [dB]},
axis background/.style={fill=white},
xmajorgrids,
xminorgrids,
ymajorgrids,
legend style={at={(0.99,0.01)}, anchor=south east, legend cell align=left, align=left, draw=white!15!black}
]
\addplot [color=mycolor1,line width=1pt,mark=star, mark options={solid, fill=white!60!mycolor1}]
  table[row sep=crcr]{%
180	16.8095286636128\\
200	16.7920584557472\\
300	16.7186630239659\\
600	16.573419306428\\
900	16.4935438831659\\
1080	16.4659248573284\\
1350	16.4253579897316\\
1620	16.3992072883036\\
1800	16.3908524382923\\
2025	16.3658612462875\\
2700	16.3231039766982\\
5400	16.2601264531003\\
};
\addlegendentry{$v=0$}

\addplot [color=mycolor2, line width=1pt, mark=pentagon*, mark options={solid, fill=white!60!mycolor2}]
  table[row sep=crcr]{%
180	16.9301826425751\\
200	16.895703491965\\
300	16.8505426742412\\
600	16.7480795946895\\
900	16.6778949991857\\
1080	16.6436322532829\\
1350	16.5950925873343\\
1620	16.5617805153422\\
1800	16.5425361312104\\
2025	16.532219334388800\\
2700	16.4970103858426\\
5400	16.3937112851264\\
};
\addlegendentry{$v=1$}

\addplot [color=mycolor3, line width=1pt, mark=square*, mark options={solid, fill=white!60!mycolor3}]
  table[row sep=crcr]{%
180	16.9720637847051\\
200	16.9680893818949\\
300	16.9161206955154\\
600	16.8259515262904\\
900	16.7714224971975\\
1080	16.7377080004158\\
1350	16.7006457383024\\
1620	16.6649071326735\\
1800	16.6388244309677\\
2025	16.6175379451473\\
2700	16.5714196597422\\
5400	16.4477052923342\\
};
\addlegendentry{$v=2$}

\addplot [color=mycolor4, line width=1pt, mark=triangle*, mark options={solid, fill=white!60!mycolor4}]
  table[row sep=crcr]{%
180	16.9879155384635\\
200	17.0014928978227\\
300	16.9629670017132\\
600	16.8800938257214\\
900	16.8044532489469\\
1080	16.7964233928515\\
1350	16.7632571187894\\
1620	16.7237292918515\\
1800	16.7056070365163\\
2025	16.6871172457971\\
2700	16.6561770182224\\
5400	16.5289569540605\\
};
\addlegendentry{$v=3$}

\addplot [color=mycolor5,line width=1pt,mark=*, mark options={solid, fill=white!60!mycolor5}]
  table[row sep=crcr]{%
180	17.0061993260011\\
200	17.0124200900081\\
300	16.9878728895811\\
600	16.8968314389618\\
900	16.851090671405\\
1080	16.8195220992094\\
1350	16.804386708856\\
1620	16.771053864935\\
1800	16.7575768077382\\
2025	16.7268815980934\\
2700	16.6891247248068\\
5400	16.5469764665976\\
};
\addlegendentry{$v=4$}

\addplot [color=mycolor1,line width=2.5pt,mark=triangle, mark options={solid, rotate=180, fill=mycolor1}]
  table[row sep=crcr]{%
  1800	16.3908524382923\\
};
\addplot [color=mycolor5,line width=2.5pt,mark=triangle, mark options={solid, fill=mycolor5}]
  table[row sep=crcr]{%
  1800	16.7575768077382\\
 };

\draw[-, very thick, dashed, draw opacity=0.9] (1350,16.804386708856) -- (6500,16.804386708856) node[] {};
\draw[-, very thick, dashed, draw opacity=0.9] (1350,16.4253579897316) -- (6500,16.4253579897316) node[] {};
\draw[<->, very thick, draw opacity=0.9,align=center] (6500,16.80)--(6500,16.43) node[midway,right, text width = 2em] {$0.38$ dB};

\end{axis}
\node[font=\Large] at (13,10) {(a)};

\begin{axis}[%
at={(mainplot.south east)},
anchor=south east,
yshift=0.7cm,
xshift=-5.3cm,
width=0.28\textwidth,
height=0.18\textwidth,
scale only axis,
xmode=log,
log ticks with fixed point,
xmin=100,
xmax=10000,
xtick={100,1000,10000},
xticklabels={$10$,$1000$,$10000$},
xminorticks=true,
xlabel style={at={(axis description cs:0.5,-0.1)},anchor=north, font=\footnotesize	\color{white!15!black}},
xlabel={Blocklength $n$},
xtick distance=1,
ymin=-9,
ymax=0,
ylabel style={at={(axis description cs:-0.22,0.45)},anchor=north, font=\footnotesize	\color{white!15!black}},
ylabel={EDI $\hat{\Psi}$ [dB]},
xmajorgrids,
xminorgrids,
ymajorgrids,
axis background/.style={fill=white},
legend style={legend cell align=left, align=left, draw=white!15!black}
]

\addplot [color=mycolor1,line width=1pt,mark=star, mark options={solid, fill=white!60!mycolor1}]
  table[row sep=crcr]{%
180	-3.0636973595673\\
200	-2.82879976897928\\
300	-2.03982956972106\\
600	-1.00437235311927\\
900	-0.864495737865003\\
1080	-0.844241071318713\\
1350	-0.646815084392287\\
1620	-0.55213203734565\\
1800	-0.675091514207156\\
2025	-0.56103872739604\\
2700	-0.570489475458478\\
5400	-0.420066786205176\\
};

\addplot [color=mycolor2, line width=1pt, mark=pentagon*, mark options={solid, fill=white!60!mycolor2}]
  table[row sep=crcr]{%
180	-5.67105181124705\\
200	-5.35084761288977\\
300	-4.52388908540388\\
600	-3.10016130474135\\
900	-2.54268448249654\\
1080	-2.29555673477982\\
1350	-2.04404029486761\\
1620	-1.86345052489963\\
1800	-1.69345266684985\\
2025	-1.73310954412152\\
2700	-1.55292001823157\\
5400	-1.00387532152798\\
};

\addplot [color=mycolor3, line width=1pt, mark=square*, mark options={solid, fill=white!60!mycolor3}]
  table[row sep=crcr]{%
180	-7.22881031702929\\
200	-7.26776689106791\\
300	-6.30235023505596\\
600	-4.53885145405928\\
900	-3.81439054797208\\
1080	-3.39385633144086\\
1350	-3.13512394253466\\
1620	-2.87930789522114\\
1800	-2.56670184862398\\
2025	-2.51899125953954\\
2700	-2.23371336421265\\
5400	-1.60847775572023\\
};

\addplot [color=mycolor4, line width=1pt, mark=triangle*, mark options={solid, fill=white!60!mycolor4}]
  table[row sep=crcr]{%
180	-7.97650524557651\\
200	-8.08750043665999\\
300	-7.45833489533211\\
600	-5.67815066826381\\
900	-4.73893577712236\\
1080	-4.43441145308593\\
1350	-3.94181972847369\\
1620	-3.6205813285773\\
1800	-3.44634460348879\\
2025	-3.21499255259541\\
2700	-2.85972278243988\\
5400	-2.01602376077474\\
};

\addplot [color=mycolor5,line width=1pt,mark=*, mark options={solid, fill=white!60!mycolor5}]
  table[row sep=crcr]{%
180	-8.35458698907518\\
200	-8.65781761369112\\
300	-8.37006426329926\\
600	-6.57478801018213\\
900	-5.61375981753902\\
1080	-5.14123558164699\\
1350	-4.64922388268374\\
1620	-4.30996783177199\\
1800	-4.04169221299712\\
2025	-3.8771554779498\\
2700	-3.28685524252768\\
5400	-2.38300446932422\\
};

\end{axis}

\end{tikzpicture}}
\hfill
\resizebox{0.48\linewidth}{!}{\definecolor{mycolor1}{rgb}{0.00000,0.44700,0.74100}%
\definecolor{mycolor2}{rgb}{0.85000,0.32500,0.09800}%
\definecolor{mycolor3}{rgb}{0.92900,0.69400,0.12500}%
\definecolor{mycolor4}{rgb}{0.49400,0.18400,0.55600}%
\definecolor{mycolor5}{rgb}{0.46600,0.67400,0.18800}%

\begin{tikzpicture}

\begin{axis}[%
width=4.521in,
height=3.566in,
at={(0.758in,0.481in)},
scale only axis,
xmode=log,
log ticks with fixed point,
xmin=100,
xmax=10000,
xlabel style={font=\color{white!15!black}},
xlabel={Blocklength $n$},
ymin=10.6,
ymax=11.05,
ylabel style={font=\color{white!15!black}},
ylabel={$\mathrm{AIR}_{\mathrm{n}}$ [bit/4D-sym]},
axis background/.style={fill=white},
xmajorgrids,
xminorgrids,
ymajorgrids,
legend style={at={(0.99,0.01)}, anchor=south east, legend cell align=left, align=left, draw=white!15!black}
]
\addplot [color=mycolor1,line width=1pt,mark=star, mark options={solid, fill=white!60!mycolor1}]
  table[row sep=crcr]{%
180	10.628690281379\\
200	10.6699022911837\\
300	10.7507317758828\\
600	10.7904832049773\\
900	10.7906963988294\\
1080	10.7955773628752\\
1350	10.7852594481854\\
1620	10.7850287439936\\
1800	10.7864793602081\\
2025	10.7761932343254\\
2700	10.761924211783\\
5400	10.7439405949495\\
};
\addlegendentry{$v=0$}

\addplot [color=mycolor2, line width=1pt, mark=pentagon*, mark options={solid, fill=white!60!mycolor2}]
  table[row sep=crcr]{%
180	10.6838660759427\\
200	10.7105801180938\\
300	10.8270043278788\\
600	10.9020835543515\\
900	10.9078255921378\\
1080	10.9065421019504\\
1350	10.8964650477773\\
1620	10.8876100953446\\
1800	10.8801135262114\\
2025	10.880731569463250\\
2700	10.8734069767658\\
5400	10.8316880774561\\
};
\addlegendentry{$v=1$}

\addplot [color=mycolor3, line width=1pt, mark=square*, mark options={solid, fill=white!60!mycolor3}]
  table[row sep=crcr]{%
180	10.6898351088195\\
200	10.7327949735789\\
300	10.8530085808027\\
600	10.9462682762433\\
900	10.9658046091607\\
1080	10.9652852706545\\
1350	10.9619155244133\\
1620	10.9518243501713\\
1800	10.9420395458467\\
2025	10.9355227251424\\
2700	10.9209783038375\\
5400	10.864498308123\\
};
\addlegendentry{$v=2$}

\addplot [color=mycolor4, line width=1pt, mark=triangle*, mark options={solid, fill=white!60!mycolor4}]
  table[row sep=crcr]{%
180	10.6833193867348\\
200	10.7341554849778\\
300	10.8715811324066\\
600	10.9741349760663\\
900	10.9833625124371\\
1080	10.999676931681\\
1350	11.0002694987682\\
1620	10.9878075254277\\
1800	10.9836299198308\\
2025	10.978867770743\\
2700	10.9746972887577\\
5400	10.9172246591473\\
};
\addlegendentry{$v=3$}

\addplot [color=mycolor5,line width=1pt,mark=*, mark options={solid, fill=white!60!mycolor5}]
  table[row sep=crcr]{%
180	10.6732195057898\\
200	10.7297474481655\\
300	10.8663863748904\\
600	10.9768146846224\\
900	11.0095400654673\\
1080	11.0115443758814\\
1350	11.0242495741238\\
1620	11.0192918231342\\
1800	11.0171898547305\\
2025	11.003173877303\\
2700	10.995247443733\\
5400	10.9283329766967\\
};
\addlegendentry{$v=4$}

\addplot [color=mycolor1,line width=2.5pt,mark=triangle, mark options={solid, rotate=180, fill=mycolor1}]
  table[row sep=crcr]{%
  1800	10.7864793602081\\
};
\addplot [color=mycolor5,line width=2.5pt,mark=triangle, mark options={solid, fill=mycolor5}]
  table[row sep=crcr]{%
  1800	11.0171898547305\\
 };

\draw[-, very thick, dashed, draw opacity=0.9] (1350,11.0242495741238) -- (200,11.0242495741238) node[] {};
\draw[-, very thick, dashed, draw opacity=0.9] (1080,10.7955773628752) -- (200,10.7955773628752) node[] {};
\draw[<->, very thick, draw opacity=0.9,align=center] (200,10.795)--(200,11.024) node[midway,left, xshift=0.2cm,text width = 5em] {$0.23$ bit/4D-sym};

\end{axis}
\node[font=\Large] at (13,10) {(b)};

\end{tikzpicture}}
\caption{Performance analysis of L-CCDM ($v=4$) and standard CCDM ($v=0$) at $1600$ km. The 256QAM symbol sequences are generated by PAS using standard CCDM or L-CCDM ($W=100$) with blocklengths $n\in\{180,200,300,600,900,1080,1350,1620,1800,2025,2700,5400\}$. (a) Effective SNR vs. blocklength $n$, where the inset shows the corresponding EDI $\hat{\Psi}$ vs. $n$. (b) AIR vs. blocklength $n$. The filled triangles in (a)--(b) for $v=0,4$ correspond to the same triangle points as shown in Fig.~\ref{figMtx}(a), (c).}
\label{fig_blkLen}
\end{figure*}

\subsection{Blocklength-dependency}
As mentioned in Sec. III, the parameters of L-CCDM should be carefully chosen to reach a favorable trade-off between the effective SNR gains and rate loss. 
Here, we consider the performance of L-CCDM for various blocklengths $n$. 
The results for each blocklength are obtained at the optimal launch power. 
Fig.~\ref{fig_blkLen}(a) shows the blocklength-dependent effective SNR at $1600$ km. 
This blocklength-dependency has been investigated for CCDM in \cite{fehenberger2019analysis,kaiquan2021EDI}. 
We show here that L-CCDM also exhibits a similar blocklength-dependent behaviour, except that the effective SNR is increased by approximately $0.38$ dB for all values of $n$ considered here. 
Larger $n$ yields lower effective SNR for both CCDM and L-CCDM, and we attribute this to the increase in the corresponding EDIs, which is self-evident in the inset figure of Fig.~\ref{fig_blkLen}(a). Since sequences with lower EDI are selected, L-CCDM generally generates higher SNR than standard CCDM, and these effective SNR gains become slightly larger in the long blocklength regime. Fig.~\ref{fig_blkLen}(b) shows that even with the additional rate loss of flipping bits in \eqref{RateLoss2}, the AIR of L-CCDM benefits from the increased NLI-tolerance with gains up to $0.23$ bit/4D-sym at the optimum blocklength $n = 1350$ with respect to that of CCDM. 
For other blocklengths, Fig.~\ref{fig_blkLen}(b) shows that the AIR gains of L-CCDM decrease (i) because for shorter blocklengths, flipping bit rate loss $v/n$ in \eqref{RateLoss2} becomes significant, and (ii) for longer blocklengths effective SNR decreases. 

\begin{figure}[]
\centering
\resizebox{1\linewidth}{!}{\definecolor{mycolor1}{rgb}{0.00000,0.44700,0.74100}%
\definecolor{mycolor2}{rgb}{0.85000,0.32500,0.09800}%
\definecolor{mycolor3}{rgb}{0.92900,0.69400,0.12500}%
\definecolor{mycolor4}{rgb}{0.49400,0.18400,0.55600}%
\definecolor{mycolor5}{rgb}{0.46600,0.67400,0.18800}%
\begin{tikzpicture}

\begin{axis}[name=mainplot,
width=4.521in,
height=3.566in,
at={(0.758in,0.481in)},
scale only axis,
xmin=800,
xmax=2400,
xlabel style={font=\color{white!15!black}},
xlabel={Transmission Distance [km]},
ymin=9,
ymax=11.5,
ylabel style={font=\color{white!15!black}},
ylabel={$\mathrm{AIR}_{\mathrm{n}}$ [bit/4D-sym]},
axis background/.style={fill=white},
title style={font=\bfseries},
xmajorgrids,
ymajorgrids,
legend style={at={(0.99,0.99)}, anchor=north east, legend cell align=left, align=left, draw=white!15!black}
]

\addplot [color=mycolor1, line width=1pt
]
  table[row sep=crcr]{%
800	12.5044391899599\\
880	12.3083061638554\\
960	12.1149497529332\\
1040	11.9249806857376\\
1120	11.7438355112238\\
1200	11.5710895363172\\
1280	11.4037505131751\\
1360	11.2405510073332\\
1440	11.0855254514132\\
1520	10.9340592738734\\
1600	10.7886711479075\\
1680	10.6477715312059\\
1760	10.5141592165348\\
1840	10.3854230859865\\
1920	10.2630678361605\\
2000	10.1440684155836\\
2080	10.0250682356019\\
2160	9.91054863905039\\
2240	9.79776845677009\\
2320	9.6906924140402\\
2400	9.5880927154701\\
};
\addlegendentry{AIR $v=0$}

\addplot [color=mycolor5, line width=1pt
]
  table[row sep=crcr]{%
800	12.6795158484532\\
880	12.497249552183\\
960	12.3166737287451\\
1040	12.1353311462607\\
1120	11.9595934098502\\
1200	11.7875631645036\\
1280	11.6220277814513\\
1360	11.4601918785659\\
1440	11.3054896138963\\
1520	11.1559525793252\\
1600	11.0113930901481\\
1680	10.8729582496408\\
1760	10.7387857220625\\
1840	10.6097686307663\\
1920	10.4846613932095\\
2000	10.3641498135954\\
2080	10.246565756687\\
2160	10.133974558104\\
2240	10.0236980356231\\
2320	9.91744344451391\\
2400	9.81244141394119\\
};
\addlegendentry{AIR $v=4$}

\addplot [color=black, line width=1pt]
  table[row sep=crcr]{%
800	12.3586421387156\\
880	12.0729029579898\\
960	11.8176017158607\\
1040	11.5701956581982\\
1120	11.3407551578361\\
1200	11.1309687269733\\
1280	10.9352238746064\\
1360	10.7480325314819\\
1440	10.5691438288553\\
1520	10.3955480272835\\
1600	10.2287542125213\\
1680	10.0730136720106\\
1760	9.91952870563246\\
1840	9.77642572667647\\
1920	9.64322465123707\\
2000	9.51602339002137\\
2080	9.39064952713975\\
2160	9.27482142300375\\
2240	9.15652904348047\\
2320	9.04460335371348\\
2400	8.94206354884899\\
};
\addlegendentry{AIR Uniform}

\addplot [color=mycolor1,dashed, line width=1pt, mark=star, mark options={solid, fill=white!60!mycolor1}
]
  table[row sep=crcr]{%
2000    9.6\\
1680	10\\
1440	10.4\\
1280	10.8\\
};
\addlegendentry{LDPC $v=0$}

\addplot [color=mycolor5, dashed,line width=1pt,mark=*, mark options={solid, fill=white!60!mycolor5}
]
  table[row sep=crcr]{%
2160    9.6\\
1840	10\\
1600	10.4\\
1360	10.8\\
};
\addlegendentry{LDPC $v=4$}

\addplot [color=black,dashed, line width=1pt, mark=square*, mark options={solid, fill=white!60!black}
]
  table[row sep=crcr]{%
960	10.6667\\
1520	9.6\\
};
\addlegendentry{LDPC Uniform}

\draw[<->, dotted,thick,draw opacity=0.9] (2000,9.6) -- (2160,9.6) node[midway,below,xshift=2pt] {160 km};
\draw[<->, dotted,very thick,draw opacity=0.9] (2000,9.6) -- (1520,9.6) node[midway,below,xshift=0pt] {480 km};

\end{axis}
\end{tikzpicture}%
}
\caption{AIR vs. transmission distance for 256QAM. The parameters of PAS using standard CCDM and L-CCDM as well as uniform 256QAM are shown in Table~\ref{PASParam}.}
\label{AIR_Reach}
\end{figure}
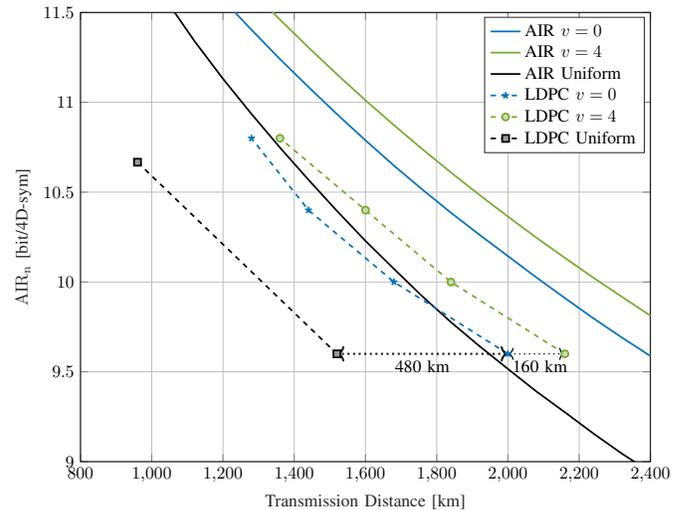

\subsection{Reach Increase}
Fig.~\ref{AIR_Reach} shows the AIR at various transmission distances for PAS using L-CCDM or CCDM (both with $n=1800$), and uniform signaling, respectively. 
Fig.~\ref{AIR_Reach} also shows the distance where the BER of uniform signaling and PAS is below $10^{-4}$ (indicated by markers). 
All of the results are obtained at the corresponding optimal launch powers. 
The transmission rate of uniform 256QAM is adjusted by changing the FEC code rate $R_c$, whilst the transmission rate of PAS is adjusted by changing shaping rate $R_s$, as can be seen in Table~\ref{PASParam}. 
At the transmission rate of $R=9.6$ bit/4D-symbol (yielding an information rate of $307$ Gbps), in addition to the $480$ km reach extension brought by standard CCDM over uniform signaling, L-CCDM with $v=4$ provides $160$ km additional transmission distance, which corresponds to a reach extension of $8\%$.

\section{Conclusions}\label{sec:Conc}
In this paper, list-encoding CCDM (L-CCDM) was proposed to mitigate some of the NLI generated during optical fiber transmission. 
L-CCDM selects NLI-tolerant channel input sequences from a set of candidate symbol sequences by measuring their energy dispersion index (EDI). 
Simulation results show that L-CCDM improves nonlinear shaping gains considerably, providing a $0.35$ dB increase in effective SNRs. The architecture of L-CCDM used in this paper can be generalized to different amplitude shapers (e.g., enumerative sphere shaping), and to different mapping strategies. In general, the architecture of L-CCDM is easy to implement, and its EDI selector module can be directly plugged into existing PAS architectures. L-CCDM showed potential for exploiting EDI in designing NLI-tolerant signaling schemes.


\ifCLASSOPTIONcaptionsoff
  \newpage
\fi

\bibliographystyle{IEEEtran}
\bibliography{NLIbib.bib}
\end{document}